\documentclass{article}

\usepackage{amsmath,amssymb}
\usepackage{epsfig,graphics}

\begin{document}

\begin{titlepage}
\vspace*{3cm}
\begin{center}
{\Large \textsf{\textbf{Negative heat capacity for a Klein-Gordon oscillator in non-commutative complex phase space}}}
\vskip 5mm
{\large \textsf{Slimane Zaim$^{*}$, Hakim Guelmamene and Yazid Delenda}}\\
\vskip 5mm
D\'{e}partement de Physique, Facult\'{e} des Sciences de la Mati\`{e}re,\\
Universit\'{e}  Batna 1, Algeria.
\vskip 4mm
{\large\textsf{\textbf{Abstract}}}
\end{center}
\begin{quote}
We obtain exact solutions to the two-dimensional Klein-Gordon oscillator in a non-commutative complex phase space to first order in the non-commutativity parameter. We derive the exact non-commutative energy levels and show that the energy levels split to $2m$ levels. We find that the non-commutativity plays the role of a magnetic field interacting automatically with the spin of a particle induced by the non-commutativity of complex phase space. The effect of the non-commutativity parameter on the thermal properties is discussed. It is found that the dependence of the heat capacity $C_V$  on the non-commutative parameter gives rise to a negative quantity. Phenomenologically, this effectively confirms the presence of the effects of self-gravitation induced by the non-commutativity of complex phase space.
\end{quote}
\vspace*{2mm}


\noindent\textbf{\sc Pacs numbers}:  02.40.Gh, 03.65.Ge, 05.
\vskip 10mm
\vspace*{20mm}
{\textsf{$^{*}$Corresponding Author, E-mail: zaim69slimane@yahoo.com}}
\end{titlepage}

\section{Introduction}

There are many papers in the literature which are devoted to the study of various aspects of the Klein-Gordon oscillator in non-commutative (NC) space and NC phase space with the usual time coordinate $\left[1-3\right]$. However the extension of this study to the case of a two-dimensional (2D) NC \emph{complex} space is limited $\left[4,5\right]$. This topic is still very interesting since its phenomenological implications are important.

This paper is organized as follows. In Section $2$, we discuss the Klein-Gordon oscillator in NC complex space. In Section $3$, we study the Klein-Gordon oscillator in NC complex phase space. Then, the thermodynamic properties are studied in section $4$. Finally, section $5$ is devoted to a discussion.

\section{2D Klein-Gordon oscillator in NC complex space}

In a complex space, the NC complex coordinate operators $\left(\hat{z},\hat{\bar{z}}\right)$ and momentum operators $\left(\hat{p}_{z},\hat{p}_{\bar{z}}\right)$ in 2D space are defined by $\left[4\right]$:
\begin{align}
\hat{z} &=\hat{x}+i\hat{y}=z+i\theta p_{\bar{z}}\,, &\hat{\bar{z}}&=\hat{x}-i\hat{y}=\bar{z}-i\theta p_{z}\,,\\
\hat{p}_{z}&=p_{z}\,, &\hat{p}_{\bar{z}}&= p_{\bar{z}}\,.
\end{align}
These operators satisfy the following commutation relations:
\begin{align}
\left[\hat{z},\hat{\bar{z}}\right] & =  2\theta\,,& \left[\hat{z},\hat{p}_{z}\right]&= \left[ \hat{\bar{z}},\hat{p}_{\bar{z}}\right]=0\,,\notag\\
\left[\hat{z},\hat{p}_{\bar{z}}\right] &= \left[ \hat{\bar{z}},\hat{p}_{z} \right] =2\hbar\, , &\left[ \hat{p}_{z},\hat{p}_{\bar{z}}\right]&=0\,.
\end{align}

Now, following ref. $[4]$, we review the Klein-Gordon oscillator in NC complex space. The Klein-Gordon oscillator in 2D complex space is defined by the following equation:
\begin{equation}
\left(2p_{\bar{z}}+im\omega z\right) \left(2p_{z}-im\omega \bar{z}\right)\psi=\left(E^2-m^2\right)\psi\,,
\end{equation}
which can be rewritten in commutative space as:
\begin{equation}
\left(p_x^2+p_y^2+m^2\omega^2\left(x^2+y^2\right)+2m\omega L_z\right)\psi =\left(E^2-m^2+2m\omega\right)\psi\,,
\end{equation}
with energy eigenvalues:
\begin{equation}
E^2=2m\omega \left(n_x+n_y+m_\ell\right)+m^2\,.
\end{equation}

In the NC complex space the Klein-Gordon oscillator is described by the following equation:
\begin{multline}
\left(
\begin{array}{cc}
\left( 2p_{z}+im\omega \hat{\bar{z}}\right)\left(2p_{\bar{z}}-im\omega \hat{z}\right) & 0 \\
0 & \left(2p_{\bar{z}}-im\omega \hat{z}\right)\left(2p_{z}+im\omega \hat{\bar{z}}\right)
\end{array}
\right) \psi =\\=\left(E^2-m^2\right) \psi\,.
\end{multline}

To solve this equation we use the NC complex coordinates:
\begin{align}
\hat{z} &= z + i \theta p_{\bar{z}}\,, \\
\hat{\bar{z}} &= \bar{z}-i\theta p_{z}\,, \\
\hat{p}_{z} &=p_{z}\,,\qquad \hat{p}_{\bar{z}}=p_{\bar{z}}\,.
\end{align}
Inserting eqs. (8)-(10) into eq. (7), we have:
\begin{multline}
\left[\left(1+\frac{m\omega \theta}{2}\right)^2\left(p_x^2+p_y^2\right)+m^2\omega^2\left(x^2+y^2\right)-2m\omega L_z-m^2\omega^2\theta \left(L_z\pm 1\right)\right]\psi\\=\left(E^2-m^2+2m\omega\right)\psi\,.
\end{multline}

The energy eigenvalues are given by:
\begin{equation}
E^2=2m\omega_\theta \left(n_x+n_y+1\right) +2m\omega_\theta\left(m_\ell \pm 1\right) +m^2\,,
\end{equation}
with $\omega_\theta=\omega (1-m\omega\theta/2)$. Such effects are similar to the normal Zeeman splitting of a particle with spin $1/2$ and thus the degeneracy of energy levels is completely removed. The oscillator is positioned in the four equivalent points $(z\uparrow ,\bar{z} \uparrow ,z\downarrow ,\bar{z}\downarrow )\Leftrightarrow (z,\bar{z},-z,- \bar{z})$. Therefore the eigenfunction $\psi(z,\bar{z})$ takes values in $C^{4}$, spin up, spin down, particle, antiparticle. This oscillator is described by two double-component spinors $[4,6]$:
\begin{equation*}
\psi_{n0}\binom{\psi_{n0}^{+}}{\psi_{n0}^{-}}\,,\qquad \text{and} \qquad \psi_{0n}\binom{\psi_{0n}^{+}}{\psi _{0n}^{-}}\,,
\end{equation*}
where the sign $(\pm)$ signifies spin up or down, and the wave functions $\psi_{n0}$ and $\psi_{0n}$ have the following form:
\begin{subequations}
\begin{align}
\psi_{n0}\left(z,\bar{z}\right) &=\sqrt{\frac{(m\omega)^{n+1}}{\pi n!}}\,z^n\exp\left(-\frac{m\omega}{2}z\bar{z}\right), \\
\psi_{0n}\left(z,\bar{z}\right) &=\sqrt{\frac{(m\omega)^{n+1}}{\pi n!}}\,\bar{z}^n\exp\left(-\frac{m\omega}{2}z\bar{z}\right).
\end{align}
\end{subequations}

\section{2D Klein-Gordon oscillator in NC complex phase space}

In a NC complex phase space we replace the coordinate and momentum operators (relations (8-10)) by:
\begin{align}
\hat{z} & = z+i\theta p_{\bar{z}}\,,\\
\hat{\bar{z}} &=\bar{z}-i\theta p_{z}\,,\\
\hat{p}_{z} &=p_{z}+i\frac{\bar{\theta}}{4}\bar{z}\,,\\
\hat{p}_{\bar{z}} &=p_{\bar{z}}-i\frac{\bar{\theta}}{4}z\,.
\end{align}
Inserting eqs. (14)-(17) into eq. (7), we have two equations:
\begin{align}
\Bigg[& \left(1-\frac{m\omega\theta}{2}\right)^2\left(p_x^2+p_y^2\right)+m^2\omega^2\left(1+\frac{\bar{\theta}}{2m\omega}\right)^2\left(x^2+y^2\right)+\notag\\
&+2m\omega L_z-m^2\omega^2\left(\theta+\frac{\bar{\theta}}{m^2\omega^2}\right) \left(L_z+1\right)\Bigg]\psi=\left(E^2-m^2+2m\omega\right)\psi\,,
\end{align}
and
\begin{align}
\Bigg[&\left(1-\frac{m\omega\theta}{2}\right)^2\left(p_x^2+p_y^2\right)+m^2\omega^2\left(1+\frac{\bar{\theta}}{2m\omega}\right)^2\left(x^2+y^2\right)+\notag\\
&+2m\omega L_z-m^2\omega^2\left(\theta+\frac{\bar{\theta}}{m^2\omega^2}\right) \left(L_z-1\right)\Bigg]\psi=\left(E^2-m^2+2m\omega\right)\psi\,.
\end{align}
The equations (18) and (19) are similar to the equation of motion for a fermion of spin $1/2$ in a constant magnetic field. Under these conditions the equations (18) and (19) take the following form:
\begin{align}
\Bigg[&\left(1+\frac{m\omega\theta}{2}\right)^2\left(p_x^2+p_y^2\right)+m^2\omega^2\left(1+\frac{\bar{\theta}}{2m\omega}\right)^2\left(x^2+y^2\right)+\notag\\
&-2m\omega L_z-m^2\omega^2\left(\theta+\frac{\bar{\theta}}{m^2\omega^2}\right)\left(L_z+2s_z\right)\Bigg]\psi=\left(E^2-m^2+2m\omega\right)\psi\,,
\end{align}
with $s_z=\pm 1/2$. The energy eigenvalues are given by:
\begin{equation}
E^2=2m\Omega_\theta \left(n_x+n_y+1\right)-2m\omega\left(m_\ell+1\right)-m^2\omega^2\left(\theta+\frac{\bar{\theta}}{m^2\omega^2}\right)\left(m_\ell\pm 1\right)+m^2\,,
\end{equation}
where
\begin{equation}
\Omega_\theta=\omega \left(1+\frac{m\omega\theta}{2}\right)\left(1+\frac{\bar{\theta}}{2m\omega }\right).
\end{equation}
We have thus shown that the non-commutativity effects are manifested in energy levels and thus the degeneracy of the levels is completely removed, so that they are split into ($2m_{\ell})$ levels, similarly to the effects of a magnetic field interacting automatically with the spin of a particle.

\section{Thermodynamic properties of the 2D Klein-Gordon oscillator in NC complex phase space}

The thermodynamic functions associated with the NC complex oscillator are also of interest. The eigenvalues of the 2D Klein-Gordon oscillator in NC complex phase space are $\left[7\right]$:
\begin{equation}
E^\pm = \pm m\sqrt{\lambda_{\theta \bar{\theta}}+\gamma_{\theta\bar{\theta}} n}\,,\qquad n=0,1,2,\hdots\,,
\end{equation}
where
\begin{equation}
\lambda_{\theta \bar{\theta}}^{\ell}=1+\gamma_{\theta\bar{\theta}}-\frac{2\omega}{m}\left(\ell+1\right)-\omega^2\left(\theta+\frac{\bar{\theta}}{m^2\omega^2}\right)\left(\ell\pm 1\right)\,,\qquad\ell=0,1,\hdots\,,
\end{equation}
and
\begin{equation}
\gamma_{\theta \bar{\theta}}=2\frac{\Omega_\theta}{m}\,.
\end{equation}

We concentrate, firstly, on the calculation of the partition function $Z(\beta,\theta,\bar{\theta})$, defined as:
\begin{equation}
Z(\beta ,\theta ,\bar{\theta})=\sum_{n,s}\exp\left[-\beta\left(E_{n,s}-E_{0,s}\right)\right],
\end{equation}
where $\beta =1/k_{B}T$ is the Boltzmann factor, and $E_0$ is the background energy correspond to $n=0$. Therefore we have the single-oscillator partition function with $\ell=0$, from eq. (26):
\begin{align}
Z(\beta ,\theta ,\bar{\theta})=&\sum_{n=0}^\infty\exp\left[-\beta m\left(\sqrt{1+\gamma_{\theta \bar{\theta}}n}-1\right)\right]+\notag\\
&+\sum_{n=0}^\infty\exp\left[-\beta m\left(\sqrt{\lambda_{\theta\bar{\theta}}+\gamma_{\theta \bar{\theta}}n}-\sqrt{\lambda_{\theta \bar{\theta}}}\right)\right],
\end{align}
where
\begin{equation}
\lambda_{\theta \bar{\theta}}=1+2\omega^2\left(\theta+\frac{\bar{\theta}}{m^2\omega^2}\right).
\end{equation}

On the other hand, the Euler-Maclaurin formula $\left[8\right]$ is:
\begin{equation}
\sum_{x=0}^\infty f(x)=\frac{1}{2}f(0) + \int_0^\infty f(x) dx - \sum_{p=1}^\infty \frac{1}{\left(2p\right)!}B_{2p}f^{\left(2p-1\right)}(0)\,,
\end{equation}
where
\begin{equation}
B_{2n}=\frac{2(2n)!}{\left(2\pi\right)^{2n}}\sum_{p=1}^{\infty }p^{-2n}\,,
\end{equation}
are the Bernoulli numbers.  Using the Euler-Maclaurin formula (eq. (29)), and after a simple calculation, the partition function in eq. (27) can be written as:
\begin{align}
Z(\beta,\theta,\bar{\theta})=& 1+\frac{2}{\gamma_{\theta \bar{\theta}}\beta^2}\left[\left(1+\beta\right)+\left(1+\beta\sqrt{\lambda_{\theta\bar{\theta}}}\right)\right]+\notag\\
&-\frac{B_2}{2}\left(e^\beta f_1^{\left(1\right)}+ e^{\beta \sqrt{\lambda_{\theta \bar{\theta}}}}f_2^{\left(1\right)}\right)+\notag \\
&-\frac{B_4}{24}\left(e^\beta f_1^{\left(3\right)}+e^{\beta\sqrt{\lambda_{\theta\bar{\theta}}}}f_2^{\left(3\right)}\right) +\hdots\,,
\end{align}
where
\begin{subequations}
\begin{align}
f_1^{\left(1\right)}&=-\frac{\gamma_{\theta\bar{\theta}}\beta m}{2}e^{-\beta }\,, \\
f_2^{\left(1\right)}&=-\frac{\gamma_{\theta\bar{\theta}}\beta m}{2\sqrt{\lambda_{\theta\bar{\theta}}}}e^{-\beta \sqrt{\lambda _{\theta\bar{\theta}}}}\,,
\end{align}
\end{subequations}
and
\begin{subequations}
\begin{align}
f_1^{\left(3\right)}&=\left[\frac{-3\beta m\left(\gamma_{\theta\bar{\theta}}\right)^3}{8}-\frac{3\beta^2m^2\left(\gamma_{\theta\bar{\theta}}\right)^3}{8}-\frac{3\beta^3m^3\left(\gamma_{\theta\bar{\theta}}\right)^3}{8}\right] e^{-\beta}\,,\\
f_2^{\left(3\right)}&=\left[\frac{-3\beta m\left(\gamma_{\theta\bar{\theta}}\right)^3}{8\left(\lambda_{\theta\bar{\theta}}\right)^{5/2}}-\frac{3\beta^2m^2\left(\gamma_{\theta\bar{\theta}}\right)^3}{8\left(\lambda_{\theta \bar{\theta}}\right)^2}-\frac{3\beta^3m^3\left(\gamma_{\theta\bar{\theta}}\right)^3}{8\left(\lambda_{\theta\bar{\theta}}\right)^{3/2}}\right] e^{-\beta \sqrt{\lambda_{\theta \bar{\theta}}}}\,,
\end{align}
\end{subequations}
with $B_{2}=1/6$ and $B_{4}=-1/30$. By replacing Eqs. (32)--(33) into Eq. (31), we obtain the partition function in NC complex phase space as:
\begin{align}
Z(\beta,\theta,\bar{\theta})=&1+\frac{2}{\gamma_{\theta\bar{\theta}}}\left(1+\sqrt{\lambda_{\theta\bar{\theta}}}\right)\beta^{-1}+\frac{4}{\gamma_{\theta\bar{\theta}}}\beta^{-2}+\notag\\
&+\frac{\gamma_{\theta\bar{\theta}}}{24}\left[\left(1+\frac{1}{\sqrt{\lambda_{\theta\bar{\theta}}}}\right)-\frac{\left(\gamma_{\theta\bar{\theta}}\right)^2}{80}\left(1+\frac{1}{\left(\lambda_{\theta\bar{\theta}}\right)^{5/2}}
\right)\right]\beta+\notag \\
&-\frac{\left(\gamma_{\theta\bar{\theta}}\right)^3}{1920}\left(1+\frac{1}{\left(\lambda_{\theta\bar{\theta}}\right)^2}\right)\beta^2 -\frac{\left(\gamma_{\theta\bar{\theta}}\right)^3}{1920}\left(1+\frac{1}{\left(\lambda_{\theta \bar{\theta}}\right)^{3/2}}\right)\beta^3\,.
\end{align}

Hence there is a characteristic temperature which divides the temperature range into two regions: $\beta \gg \beta_0=1/mc^2$ for very low temperatures, and $\beta \ll \beta _0$ for    very high temperatures. In
this context, we derive the thermodynamic properties of our system, such as the total energy, entropy, free energy and specific heat, which are given by:
\begin{subequations}
\begin{align}
\left\langle E(\beta,\theta,\bar{\theta})\right\rangle &= -\frac{\partial}{\partial\beta}\ln Z(\beta,\theta,\bar{\theta})\,,& C_V&=-k_B\beta^2\frac{\partial\left\langle E(\beta,\theta,\bar{\theta})\right\rangle}{\partial \beta}\,,\\
F&=-\frac{1}{\beta}\ln Z(\beta,\theta,\bar{\theta})\,,&S&=-\frac{1}{T} F-k_B\frac{\partial}{\partial \beta}\ln Z(\beta,\theta,\bar{\theta})\,.
\end{align}
\end{subequations}

\subsection{Results and discussions}

For very low temperatures the partition function in eq. (34) can be written as:
\begin{align}
Z(\beta,\theta,\bar{\theta}) &\simeq \frac{\gamma_{\theta\bar{\theta}}}{24}\left[\left(1+\frac{1}{\sqrt{\lambda_{\theta\bar{\theta}}}}\right)-\frac{\left(\gamma_{\theta\bar{\theta}}\right)^2}{80}\left(1+\frac{1}{
\left(\lambda_{\theta\bar{\theta}}\right)^{5/2}}\right)\right]\beta+\notag\\
&-\frac{\left(\gamma_{\theta\bar{\theta}}\right)^3}{1920}\left(1+\frac{1}{\left(\lambda_{\theta\bar{\theta}}\right)^2}\right)\beta^2-\frac{\left(\gamma_{\theta\bar{\theta}}\right)^3}{1920}\left(1+\frac{1}{\left(\lambda_{\theta \bar{\theta}}\right)^{3/2}}\right)\beta^3\,.
\end{align}
The mean energy $\left\langle E(\beta,\theta,\bar{\theta})\right\rangle$ of the systems is
\begin{equation}
\left\langle E(\beta,\theta,\bar{\theta})\right\rangle =-\frac{\partial}{\partial \beta}\ln Z(\beta,\theta,\bar{\theta})\sim 0\,,
\end{equation}
and the specific heat $C_V$ of the systems is:
\begin{equation}
C_V=-k_B\beta^2\frac{\partial \left\langle E(\beta,\theta,\bar{\theta})\right\rangle}{\partial \beta}\sim -3k_B\,,
\end{equation}
which is clearly a negative quantity. In other words this means that the Klein-Gordon oscillator in NC complex phase space is similar to a self-gravitating system that is discussed in the framework of a non-extensive kinetic theory (see $[9]$ and references therein).

The free energy $F$ is given by:
\begin{equation}
F=-\frac{1}{\beta}\ln Z(\beta,\theta,\bar{\theta})\sim 0\,,
\end{equation}
and the entropy of the systems is:
\begin{equation}
S=-\frac{1}{T}F-k_{B}\frac{\partial}{\partial \beta}\ln Z(\beta,\theta ,\bar{\theta})\sim 0\,.
\end{equation}

For very high temperatures the partition function in eq. (34) can be written as:
\begin{equation}
Z(\beta,\theta ,\bar{\theta})\simeq 1+\frac{2}{\gamma_{\theta \bar{\theta}}}\left(1+\sqrt{\lambda_{\theta\bar{\theta}}}\right) \beta^{-1}+\frac{4}{\gamma_{\theta\bar{\theta}}}\beta^{-2}\,,
\end{equation}
and the mean energy $\left\langle E(\beta,\theta,\bar{\theta})\right\rangle$, and the specific heat $C_V$ of the systems are given by:
\begin{align}
\left\langle E(\beta,\theta,\bar{\theta})\right\rangle &= -\frac{\partial}{\partial \beta}\ln Z(\beta,\theta,\bar{\theta})\sim 2\beta^{-1}\,,\\
C_V &=\frac{\partial\left\langle E(\beta,\theta,\bar{\theta})\right\rangle}{\partial T}\sim 2k_{B}\,.
\end{align}
Note that these results are similar to those of the Dirac oscillator under a magnetic field in a NC space $\left[5\right]$.

\begin{figure}[ht]
\centering
\includegraphics[width=0.64\textwidth]{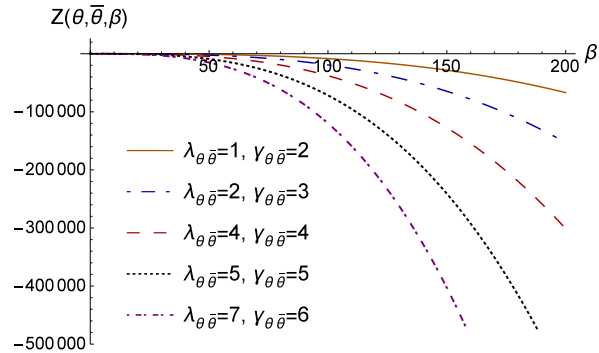}
\caption{\label{fig:Z}Comparison of the partition function $Z$ as a function of $\beta$ for different values of $\theta$ and $\bar{\theta}$.}
\end{figure}

\begin{figure}[ht]
\centering
\includegraphics[width=0.64\textwidth]{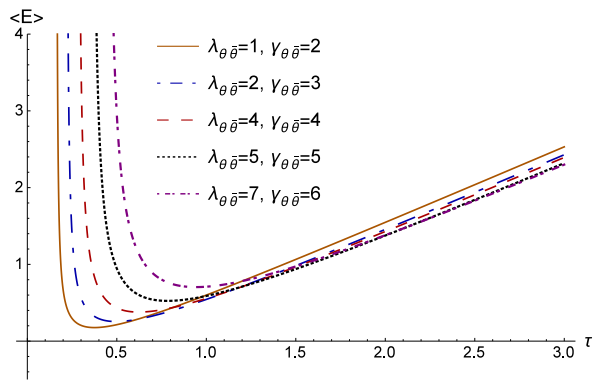}
\caption{\label{fig:E}Comparison of the thermodynamic function $E/mc^2$ of the Klein Gordon Bosons as a function of $\tau$ for different values of $\theta$ and $\bar{\theta}$.}
\end{figure}

\begin{figure}[ht]
\centering
\includegraphics[width=0.64\textwidth]{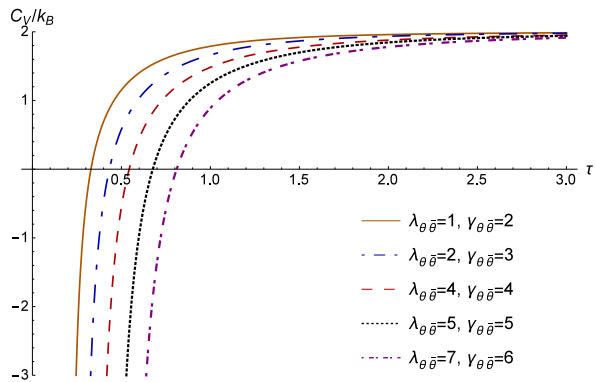}
\caption{\label{fig:Cv}Comparison of the heat capacity $C_V/k_B$ of Klein-Gordon Bosons as a function of $\tau$ for different values of $\theta$ and $\bar{\theta}$.}
\end{figure}

We shown in figures \ref{fig:Z}, \ref{fig:E} and \ref{fig:Cv} comparisons of the partition function $Z$ as a function of $\beta$, the thermodynamic function $E/mc^2$ of the Klein Gordon Bosons as a function of $\tau$ and  the heat capacity $C_V/k_B$ of Klein-Gordon Bosons as a function of $\tau$, for different values of $\theta$ and $\bar{\theta}$.

\section{Conclusions}

In this paper we started from a Klein-Gordon oscillator in a NC complex phase space. Using the Moyal product method, we derived the deformed Klein-Gordon oscillator and showed that it similar to the Klein-Gordon
equation for a particle with spin $1/2$ in a uniform magnetic field in NC phase space $\left[ 2\right]$. We solved this equation exactly and found that the NC energy levels split into $2m_{\ell}$ levels. Thus the system without spin in a NC complex coordinate space has an added advantage that the spin effect is automatically manifested. The statistical quantities of the 2D Klein-Gordon oscillator in a NC complex phase space were investigated and the effect of the NC parameter on thermal properties was discussed. It was found that the dependence of the specific heat $C_V$ on the NC parameter gives rise to a negative quantity. Phenomenologically, this effectively
confirms the presence the effects of self-gravitation at this level.

\end{document}